# Real Differences between OT and CRDT under a General Transformation Framework for Consistency Maintenance in Co-Editors

CHENGZHENG SUN, Nanyang Technological University, Singapore
DAVID SUN, Codox Inc., United States
AGUSTINA NG, Nanyang Technological University, Singapore
WEIWEI CAI, Nanyang Technological University, Singapore
BRYDEN CHO, Nanyang Technological University, Singapore

OT (Operational Transformation) was invented for supporting real-time co-editors in the late 1980s and has evolved to become a collection of core techniques widely used in today's working co-editors and adopted in major industrial products. CRDT (Commutative Replicated Data Type) for co-editors was first proposed around 2006, under the name of WOOT (WithOut Operational Transformation). Follow-up CRDT variations are commonly labeled as "*post*-OT" techniques capable of making concurrent operations *natively* commutative in co-editors. On top of that, CRDT solutions have made broad claims of superiority over OT solutions, and routinely portrayed OT as an incorrect, complex and inefficient technique. Over one decade later, however, CRDT is rarely found in working co-editors, and OT remains the choice for building the vast majority of today's co-editors. Contradictions between the reality and CRDT's purported advantages have been the source of much confusion and debate in co-editing research and developer communities. Have the vast majority of co-editors been unfortunate in choosing the faulty and inferior OT, or those CRDT claims are false? What are the real differences between OT and CRDT for co-editors? What are the key factors and underlying reasons behind the choices between OT and CRDT in the real world? A thorough examination of these questions is relevant not only to researchers who are exploring the frontiers of co-editing technologies and systems, but also to practitioners who are seeking viable techniques to build real world applications. To seek truth from facts, we set out to conduct a comprehensive and critical review on representative OT and CRDT solutions and working co-editors based on them. From this work, we have made important discoveries about OT and CRDT, and revealed facts and evidences that refute CRDT claims over OT on all accounts. We report our discoveries in a series of articles and the current article is the first one in this series.

In this paper, we present a general transformation framework for consistency maintenance in co-editors, which was distilled from dissecting and examining representative OT and CRDT solutions (and other alternative solutions) during this work, and report our discoveries under the guidance of this framework. In particular, we reveal that CRDT is like OT in following a general transformation approach, but achieves the same transformation *indirectly*, in contrast to OT *direct* transformation approach; and CRDT is not natively commutative for concurrent co-editing operations, but has to achieve the same OT commutativity indirectly as well, with consequential correctness and complexity issues. Uncovering the hidden transformation nature and demystifying the commutativity property of CRDT provides much-needed clarity about what CRDT really *is* and *is not* to co-editing, and serves as the foundation to explore the real differences between OT and CRDT in correctness, complexity, implementation, and real world applications, which are reported in follow-up articles. We hope discoveries from this work help clear up common misconceptions and confusions surrounding OT and CRDT, and accelerate progress in co-editing technology for real world applications.

**KEYWORDS**

Operational Transformation (OT); Commutative Replicated Data Type (CRDT); Concurrency Control; Consistency Maintenance; Real-Time Collaborative Editing; Distributed/Internet/Cloud Computing Technologies and Systems; Computer Supported Cooperative Work (CSCW) and Social Computing.

**Corresponding author**: Chengzheng Sun, School of Computer Science and Technology, Nanyang Technological University, Singapore. Email: CZSun@ntu.edu.sg; URL: https://www.ntu.edu.sg/home/CZSun

## 1 INTRODUCTION

Real-time co-editors allow multiple geographically dispersed people to edit shared documents at the same time and see each other's updates instantly [1,6,14-17,39,44,55,56,61,73,79]. One major challenge in building such systems is consistency maintenance of shared documents in the face of concurrent editing, under high communication latency networks like the Internet, and without imposing interaction restrictions on human users [14,55,56].

Operational Transformation (OT) was invented to address this challenge [14,55,62,73] in the late 1980s. OT introduced a framework of transformation algorithms and functions to ensure consistency in the presence of concurrent user activities. The OT framework is grounded in established distributed computing theories and concepts, principally in *concurrency* and *context* theories [25,55,67,68,84]. Since its inception, the scope of OT research has evolved from the initial focus on consistency maintenance (or concurrency control) to include a range of key collaboration-enabling capabilities, including *group undo* [39,45,58,59,67,68], and *workspace awareness* [1,20,61]. In the past decade, a main impetus to OT research has been to move beyond plain-text co-editing [6,14,21,39,44,55,56,59,63,71,72,78], and to support rich-text co-editing in word processors [61,66,69,83], HTML/XML Web document co-editing [11], spreadsheet co-editing [70], 3D model co-editing in digital media design tools [1,2], and file synchronization in cloud storage systems [3]. OT-based co-editors have also evolved from allowing people to use the same editors in one session (*homogeneous* co-editing) [12,56,61,75], to supporting people to use different editors in the same session (*heterogeneous* co-editing) [9]. Recent years have seen OT being widely adopted in industry products as the core collaboration-enabling technique, ranging from battle-tested online collaborative rich-text editors like Google Docs[1][12], to emerging start-up products, such as Codox Apps[2].

In addition to OT, a variety of alternative techniques for consistency maintenance in co-editors had been explored in the past decades [15,17,19,42,43,73]. One notable class of techniques is CRDT[3] (Commutative Replicated Data Type) for co-editors [4,5,8,26,33,38,40-42,46,48,49,80-82]. The first CRDT solution appeared around 2006 [41,42], under the name of WOOT (WithOut Operational Transformation). One motivation behind WOOT was to solve the FT (*False Tie*) puzzle in OT for plain-text co-editors [54,56], using a radically different approach from OT. Since then, numerous WOOT revisions (e.g. WOOTO [81], WOOTH [4]) and alternative CRDT solutions (e.g. RGA [46], Logoot [80,82], LogootSplit [5]) have appeared. In CRDT literature, CRDT has commonly been labelled as a "*post*-OT" technique that makes concurrent operations *natively* commutative, and does the job "*without operational transformation*" [41,42], and even "*without concurrency control*" [26]. CRDT solutions have made broad claims of superiority over OT solutions, and routinely portrayed OT as an incorrect and inferior technique (see footnote 4).

After over one decade, however, CRDT is rarely found in working co-editors or industry co-editing products, and OT remains the choice for building the vast majority of today's co-editors. The contradictions between the reality and CRDT's purported advantages have been the source of much confusion and debate in co-editing research and developer communities[4]. Have the majority of existing co-editors been unfortunate in choosing the faulty and inferior OT, or those CRDT claims over OT are false? What are the real differences between OT and CRDT for co-editors in terms of their basic approaches, correctness, complexity, and implementation? What are the key factors and underlying reasons behind the choices between OT and CRDT in the real world? We believe a thorough examination of these questions is relevant not only to researchers exploring the frontiers of collaboration technologies and systems, but also to practitioners seeking viable techniques to build real world collaboration tools and applications.

[1] https://www.google.com/docs/about/

[2] https://www.codox.io

[3] In literature, CRDT can refer to a number of different data types [49]. In this paper, we focus *exclusively* on CRDTs for *text co-editors*, which we abbreviate as "CRDT" in the rest of the paper, though occasionally we use "CRDT for co-editors" for emphasizing this point and avoiding misinterpretation.

[4] We posted an early version of our report on this work at https://arxiv.org/abs/1810.02137, in Octo. 2018, which attracted wide interests and discussions in public blogs (among academics and practitioners) and private communications (between readers and authors). This link, at https://news.ycombinator.com/item?id=18191867, hosts some representative comments

To seek truth from facts, we set out to conduct a comprehensive and critical review on representative OT and CRDT solutions and working co-editors based on them, which are available in publications and/or from publicly accessible open-source project repositories. We explore *what, how,* and *why* OT and CRDT solutions are different and the consequences of their differences from both an algorithmic angle and a system perspective. We know of no existing work that has made similar attempts. In this work, we focus on OT and CRDT solutions to consistency maintenance in *real-time* co-editing, as it is the foundation for other co-editing capabilities, such as group undo and issues related to *non-real-time* co-editing, which will be covered in future work.

The topics covered in this work are complex, diverse and comprehensive, and the bulk of outcomes from this work are well beyond the scope of a single conference paper. To cope with the complexity and diversity of topics and readerships, and take into account of feedback to a prior version of our report on this work (see footnote 4), we have organized the outcome materials of this work in a series of three *related* but *self-contained* papers, including the current paper and two follow-ups [74,75]. We briefly describe the main results reported in these three papers below.

In the current paper, we present a general transformation framework, together with major discoveries about OT and CRDT under this framework. The general framework provides a common ground for describing, examining and comparing a variety of consistency maintenance solutions in co-editing, including, but not limited to, OT and CRDT. With the guidance of this framework, we have made important discoveries about OT and CRDT, some of which are quite surprising. For instance, we found that CRDT is like OT in following a general transformation approach, but achieves the same transformation *indirectly*, rather than *directly* as OT does. Moreover, we found that CRDT is not natively commutative for concurrent operations in co-editors, as often claimed (a myth), but has to achieve the same OT commutativity indirectly as well. Uncovering the hidden transformation nature and demystifying the commutativity property of CRDT provides much-needed clarity about what CRDT really *is* and *is not* to co-editing, which serves as the foundation to reveal real differences between OT and CRDT for co-editors in correctness and complexity, as well as in building real world co-editors, reported in [74,75]. Materials in the current paper are presented at high levels and require no in-depth co-editing technical background from readers; advanced knowledge in co-editing is nevertheless beneficial to gain deep understanding of the new perspectives and insights on various co-editing issues presented in this paper.

---

and opinions on various co-editing issues addressed in our article. One well-known CRDT advocate commented there: *"The argument of Sun's paper seems to be that CRDTs have hidden performance costs. Perhaps this is true. This completely misses the main point. OT is complex, the theory is weak, and most OT algorithms have been proven incorrect (...). AFAIK, the only OT algorithm proved correct is TTF, which is actually a CRDT in disguise. In contrast, the logic of CRDTs is simple and obvious. We know exactly why CRDTs converge. ... Disclaimer: I did not read the paper in detail, just skimmed over it."* This CRDT advocate basically re-iterated some common CRDT claims against OT, which re-confirms the liveness of ongoing debates, and warrants a thorough examination of those CRDT claims. Without reading the paper in detail, this CRDT advocate clearly missed the facts and arguments presented in our paper. In fact, we had examined all points mentioned above (and beyond), and revealed facts and evidences that refute those CRDT claims on all accounts. Readers may make independent judgement after reading our papers in this series.

Built on the results reported in the first paper, we proceed to examine *real differences between OT and CRDT in correctness and complexity for consistency maintenance in co-editors* in [74]. We dissect representative OT and CRDT solutions, and explore how different basic approaches to realizing the same general transformation, i.e. the *direct* and *concurrency-centric* transformation approach taken by OT, and the *indirect* and *content-centric* transformation approach taken by CRDT, had led to different technical issues and challenges, and consequential correctness and complexity problems and solutions. Furthermore, we reveal hidden complexity issues and algorithmic flaws with representative CRDTs, and discuss common myths and facts related to correctness, time and space complexity, and simplicity of OT and CRDT solutions. Materials in this paper are technical in nature, so in-depth understanding of the technical contents and their implications require advanced co-editing background from readers.

Furthermore, we examine *real differences between OT and CRDT in building co-editing systems and real world applications* in [75]. In particular, we investigate the role of building working co-editors in shaping OT and CRDT research and solutions, and the consequential differences in the practical application and real world adoption of OT and CRDT. In this paper, we review the evolution of co-editors from research vehicles to real world applications, and discuss representative OT-based co-editors and alternative approaches in industry products and open source projects. Moreover, we evaluate CRDT-based co-editors in relation to published CRDT solutions, and clarify myths surrounding system implementation and "peer-to-peer" co-editing. Materials in [75] should be of particular interest to researchers investigating co-editing system technologies and practitioners seeking viable techniques for building real world applications.

In summary, this series of three papers present our discoveries about OT and CRDT, with respect to their basic approaches to consistency maintenance, correctness, complexity, implementation, and real world applications. This work has revealed facts and evidences that refute CRDT superiority claims over OT on all accounts, which helps to explain the underlying reasons behind the choices between OT and CRDT in real world co-editors. These results are relevant to both researchers and practitioners in the co-editing community. For researchers, these results can help them to better understand the start-of-the-art in the frontiers of co-editing, learn from the experiences and lessons of prior OT and CRDT work, and avoid being trapped in pitfalls or irrelevant issues, which lead to nowhere. For practitioners, these results can help them to learn which published solutions really work or do not, choose viable techniques to build real world applications, and avoid being misled by false claims in literature and wasting time and resources.

The rest of the current paper is organized as follows. In Section 2, we present the basic ideas of a general transformation approach. In Section 3, we examine the basic approaches taken by OT and CRDT to realize the general transformation approach. In Section 4, we present the general transformation framework, describe OT and CRDT under this framework, and reveal the differences between the OT direct and CRDT indirect transformation approaches. Furthermore, we describe the TTF (Tombstone Transformation Function [43]) solution – a hybrid of CRDT and OT – under this framework, and demonstrate the generality of the framework. In Section 5, we further explore the hidden transformation nature of CRDT, clear up common myths and misconceptions about the commutativity property of CRDT, and reveal general differences between OT and CRDT in time and space complexity without diving into details of specific algorithms. In Section 6, we summarize the major results in this paper.

## 2 BASIC IDEAS OF A GENERAL TRANSFORMATION APPROACH

Modern real-time co-editors have commonly adopted a replicated architecture: the editor application and shared documents are replicated at all co-editing sites. A user may directly edit the local document replica and see the edit effect immediately; local edits are promptly propagated to remote sites for real-time replay there.

There are two basic ways to propagate local edits: one is to propagate edits as *operations* [14,42,55,56,80]; the other is to propagate edits as *states* [15]. Most real-time co-editors, including those based on OT and CRDT, have adopted the operation propagation approach for communication efficiency, among others. In the rest of this article, the operation approach is assumed for all editors discussed.

A central issue shared by all co-editors is: how an operation generated from one replica can be replayed at other replicas, in the face of concurrent operations, to achieve consistent results across all replicas. Co-editors are generally required to meet three consistency requirements [56]: the first is *causality-preservation*, i.e. operations must be executed in their causal-effect orders, as defined by the happen-before relation [25]; the second is *convergence*, i.e. replicas must be the same after executing the same collection of operations; and the third is *intention-preservation*, i.e. the effect of an operation on the local replica from which this operation was generated must be preserved at all remote replicas in the face of concurrency.

A general approach to achieving both convergence and intention-preservation, invented in past co-editing research, is based on the notion of *transformation*, i.e. an original operation is transformed (one way or another) into a new version, according to the impact of concurrent operations, so that executing the new version on a remote replica can achieve the same effects as executing the original operation on its local replica [56]. This approach allows concurrent operations to be executed in different orders (thus being *commutative*), but in *non-original* forms[5]. Causality-preservation can be achieved by adopting suitable distributed computing techniques [14,25,55], without involving the aforementioned transformation.

The transformation approach can be illustrated by using a real-time plain text co-editing scenario in Fig. 1-(a). The initial document state *"abe"* is replicated at two sites. Under the transformation consistency maintenance scheme, users may freely edit replicated documents to generate operations. Two operations, $O_1$ = *D*(1) (to delete the character at position 1) and $O_2$ = *I*(*2,"c"*) (to insert character *c* at position 2), are generated by User A and User B, respectively. These two operations are concurrent as they are generated without the knowledge of each other [25,55]. The two operations are executed *as-is* immediately at local sites to produce *"ae"* and "*abce*", respectively; then propagated to remote sites for replay.

If there was no any consistency maintenance scheme in a co-editor, the two operations would be executed in their original forms and in different orders at the two sites, due to network communication delay. This would result in inconsistent states "*aec*" (under the *shadowed cross* at User A) and "*ace*" (at User B), as shown in Fig. 1-(a). Under the transformation-based consistency maintenance, however, a co-editor may execute a remote operation in a transformed form that takes into account the impact of concurrent operations, or *concurrency-impact* in short.

In this example, the two concurrent operations are executed as follows:

- At User A, $O_1$ has left-shifting concurrency-impact on $O_2$. The transformation scheme creates a new $O_2'$ = *I*(*1,"c"*) from the original $O_2$ = *I*(*2, "c"*), to insert "c" at position 1.
- At User B, $O_2$ has no shifting concurrency-impact on $O_1$. So, the original $O_1 = O_1'$ = *D*(1) can be applied to delete "b" at position 1.

Executing $O_2'$ at User A and $O_1'$ at User B, respectively, would result in the same document state "ace", which is not only convergent, but also preserves the original effects of both $O_1$ and $O_2$, thus meeting the intention-preservation requirement as well [55,56]. We draw attention to the fact that $O_1$ and $O_2$ are executed in different orders at two sites but achieve the same result,

[5] In contrast, an alternative approach, called *serialization*, forces all operations to be executed in *the same order* and in *original forms* [14,17,55]. It has been shown serialization is unable to achieve intention-preservation [56].

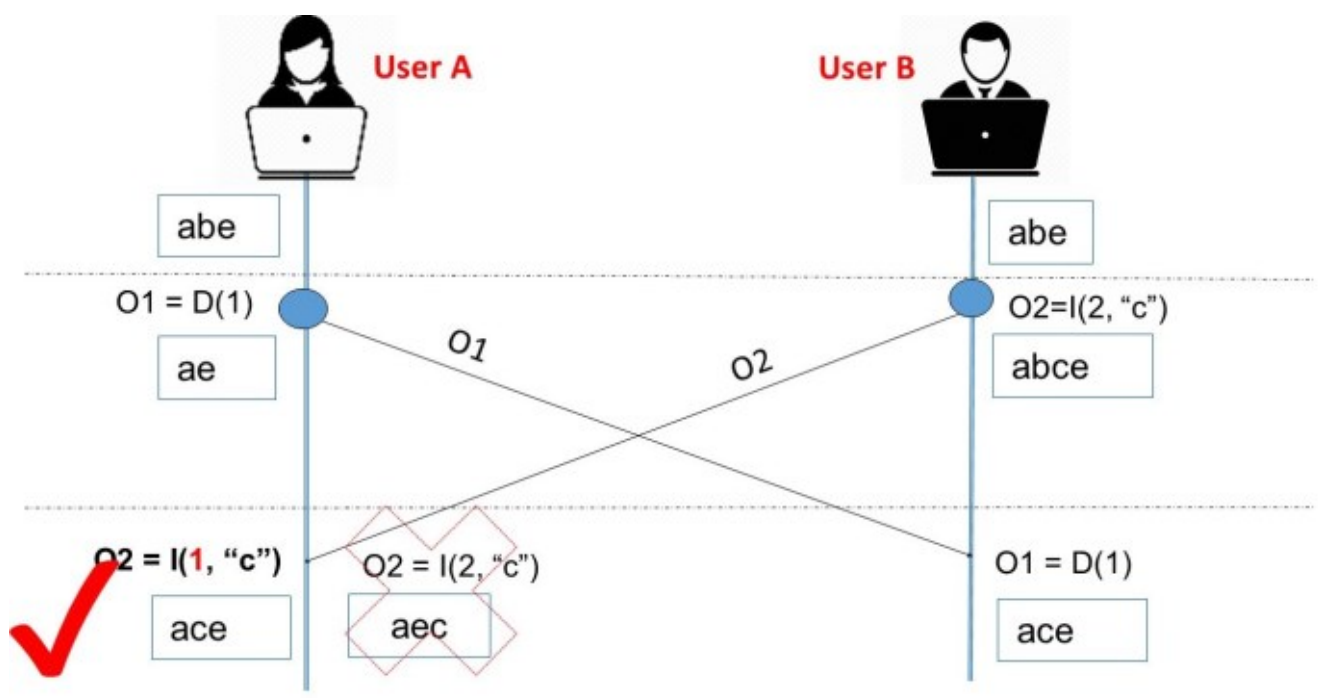

(a) Basic idea of the general transformation, which allows concurrent operations to be executed in different orders but achieve the same result, i.e. making concurrent operations *commutative* on replicated documents in real-time co-editors.

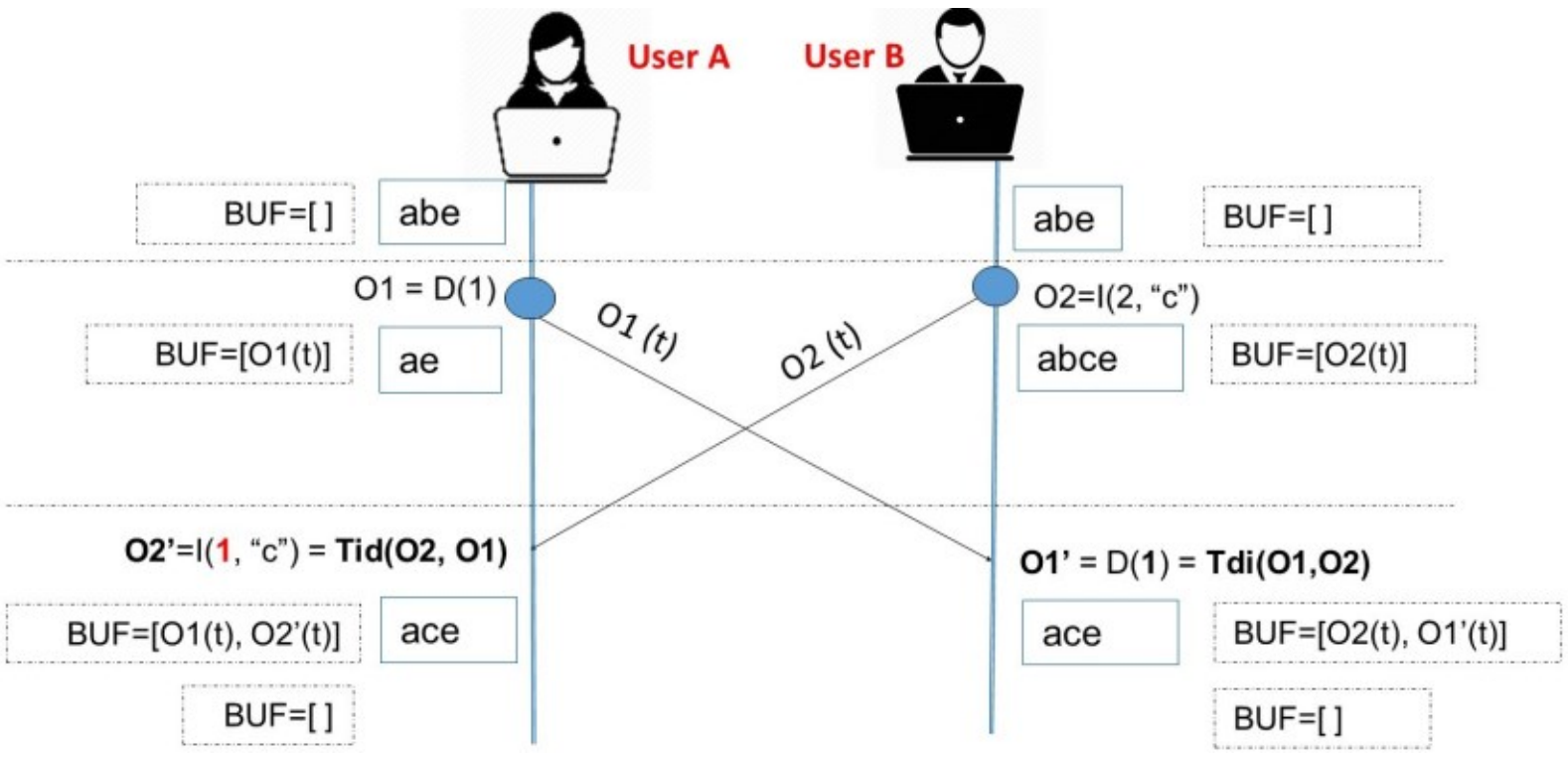

(b) The OT approach to realizing the general transformation (elaborated in Section 3.1.2). OT propagates *position*-based operations defined on the external document state.

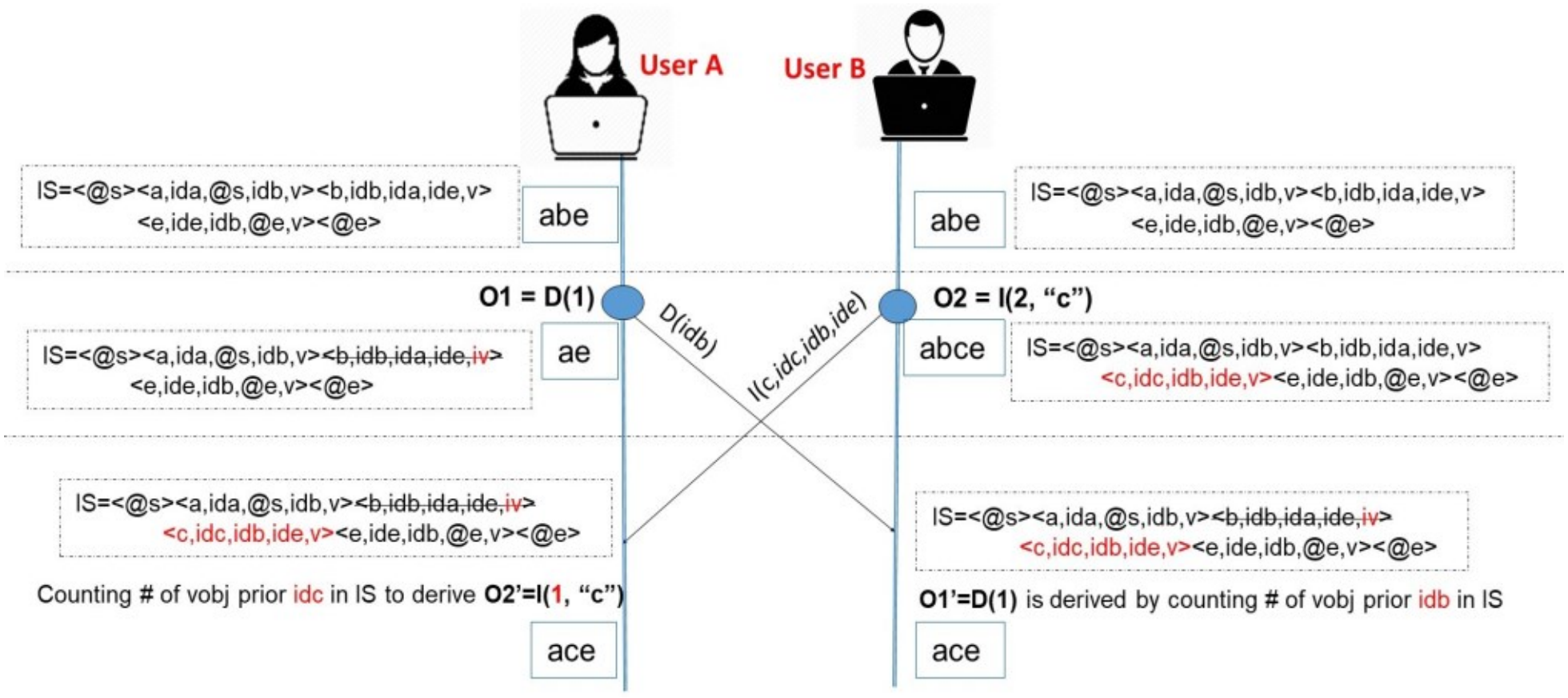

(c) The WOOT approach to realizing the general transformation (elaborated in Section 3.2.2). WOOT propagates *identifier*-based operations defined on the internal object sequence.

Fig. 1 Illustrating OT and CRDT different approaches to realizing the same general transformation.

as seen in Fig. 1-(a), which illustrates that the transformation approach has the capability of making concurrent operations commutative on replicated documents.

The consistency maintenance problem and solution illustrated in Fig. 1-(a) should look familiar to readers with some background in OT. Indeed it has often been used to explain basic OT ideas [14,55,56,71,72]. What might be surprising to many is that the same formulation of problem and solution apply equally to CRDT as well: CRDT was proposed to address the same consistency maintenance issues in co-editors, and has actually followed the same transformation approach as well. We illustrate this point by instantiating the same example scenario under the general transformation approach in Fig. 1-(a), with OT and CRDT specific realizations in Fig. 1-(b) and (c), respectively. Detailed elaborations of inner workings of OT and CRDT under this example scenario are provided in Sections 3.1 and 3.2, respectively.

OT has been known for its very capability of *making concurrent operations commutative among replicated documents* long before CRDT appeared. What has been mysterious to many is the notion that CRDT achieves commutativity of concurrent operations *natively* or *by design,* whereas OT achieves commutativity *after the fact* [48,49]. In this work (Sections 4 and 5), we demystify this CRDT native commutativity and reveal CRDT has to achieve the same OT commutativity *after the fact* as well, albeit *indirectly,* with consequential correctness and complexity issues.

To summarize, the real differences between OT and CRDT lie not in whether their commutativity capability is native or not, but in their radically different ways of achieving the same non-native commutativity for co-editors under the same general transformation framework, which are examined in detail in the rest of this paper and follow-up papers in this series.

## 3 DIFFERENT APPROACHES TO REALIZING THE SAME TRANSFORMATION

In the next two subsections, we present the basic functional components of OT and CRDT for co-editors, and use the same co-editing scenario in Fig. 1-(a) to illustrate how OT and CRDT realize the same general transformation. Rather than reviewing individual algorithmic elements in isolation, we take a systematic and end-to-end perspective, i.e. examining the whole process from the point when an operation is generated from a local editor by a user, all the way to the point when this operation is replayed in a remote editor seen by another user. We give step by step illustrations of the general process of handling an operation at both local and remote sites under both approaches, so that the subtle but key differences between OT and CRDT can be contrasted (*the devil is in the details*).

### 3.1 The OT Approach

*3.1.1 Key Ideas and Components.* An OT solution for consistency maintenance typically consists of two key components[6]: generic *control algorithms* for managing the transformation process; and application-specific *transformation functions* for performing the actual transformation (or manipulation) on concrete operations. At each collaborating site, OT control algorithms maintain an operation buffer for saving operations that have been executed and may be *concurrent* with future operations.

The life cycle of a user-generated operation in an OT-based co-editor is sketched below.

[6] In this work, we focus *exclusively* on OT solutions that separate *generic control algorithms* from *application-specific transformation functions* [1-3,6,11,12,14,16,21,27,32,35,37,39,44,45,50-75,77-79,83-85], as they represent the majority and mainstream OT solutions, on which existing OT-based co-editors are built. In co-editing literature, however, there are other OT solutions (e.g. [28-31,47]), in which control procedures are not generic but dependent on specific types of operation and data, and transformation procedures may examine concurrency relationships among other operations as well. In those OT solutions, "*control procedure and transformation functions are not separated as in previous works – instead, they work synergistically in ensuring correctness*"[31], and different correctness criteria were used as well [28-31,47,62].

- When an operation is generated by a user at a collaborating site, this operation is immediately executed on the local document state visible to the user. Then, this operation is timestamped to capture its concurrency relationship with other operations and saved in the local buffer. Next, the timestamped operation is propagated to remote sites via a communication network.
- When an operation arrives at a remote site, it is accepted according to the causality-based condition [14,25,55]. Then, *control algorithms* are invoked to select suitable concurrent operations from the buffer, and *transformation functions* are invoked to transform the remote operation against those selected concurrent operations to produce a transformed operation (a version of the remote operation is also saved in the buffer). Finally, the transformed operation is replayed on the document visible to the remote user.

For a plain-text co-editor with a pair of *insert* and *delete* operations, a total of four transformation functions, denoted as T*ii*, T*id*, T*di*, and T*dd,* are needed for four different operation type combinations [55,62,71,72]. Each function takes two operations, compares their positional relations (e.g. *left*, *right*, or *overlapping*) to derive their concurrency impacts on each other, and adjusts the parameters of the affected operation accordingly. When extending an OT solution to editors with different data and operation models, transformation functions need to be re-designed, but generic control algorithms need no change.

*3.1.2 A Working Example for OT.* In Fig. 1-(b), we illustrate how the key components of an OT solution work together to achieve the consistent result in Fig. 1-(a). Each co-editing site is initialized with the same *external* document state *"abe"*, and an empty *internal* buffer BUF.

*Local Operations Handling.* User A interacts with the external state to generate $O_1 = D(1)$, which results in a new state "ae". Internally, the OT solution at User A would do the following:

1. Timestamp $O_1$ to produce an internal operation $O_1(t)$.
2. Save $O_1(t)$ in BUF = [$O_1(t)$].
3. Propagate $O_1(t)$ to the remote site.

Concurrently, User B interacts with the external state to generate $O_2 = I(2, \text{"}c\text{"})$, which results in a new state "*abce*". Internally, the OT solution at User B would do the following:

1. Timestamp $O_2$ to produce an internal operation $O_2(t)$.
2. Save $O_2(t)$ in BUF = [$O_2(t)$].
3. Propagate $O_2(t)$ to the remote site.

*Communication and Operation Propagation*: The basic OT approach described here is independent of specific communication structures or protocols (more elaboration on this point later). What is noteworthy here is that under the OT approach, operations propagated among co-editing sites are *position*-based operations defined on the external document state.

*Remote Operation Handling.* When $O_2(t)$ arrives at User A, OT would do the following:

1. Accept $O_2(t)$ for processing under certain conditions (e.g. causal ordering [25]).
2. Transform $O_2(t)$ into $O_2'(t)$ by:
   a. invoking the control algorithm to get $O_1(t)$ from BUF, which is concurrent and defined on the same initial document state with $O_2(t)$; and
   b. invoking the transformation function $Tid(O_2, O_1)$ to produce a transformed operation $O_2' = I(1, \text{"}c\text{"})$. The *Tid* function works by comparing the position parameters 2 and 1 in $O_2$ and $O_1$, respectively, and derives that $O_2$ is performed on the right of $O_1$ in the linear document state, and hence adjusts $O_2$ position from 2 to 1 to compensate the left shifting effect of $O_1$.
3. Save $O_2'(t)$ in BUF = [$O_1(t)$, $O_2'(t)$].
4. Replay $O_2' = I(1, \text{"}c\text{"})$ on "ae" to produce "ace".

When $O_1(t)$ arrives at User B, OT would do the following:

1. Accept $O_1(t)$ for processing under certain conditions (e.g. causal ordering [25]).
2. Transform $O_1(t)$ into $O_1'(t)$ by:
   a. invoking the control algorithm to get $O_2(t)$ from BUF, which is concurrent and defined on the same initial document state with $O_1(t)$; and
   b. invoking the transformation function $Tdi(O_1, O_2)$ to produce a new operation $O_1' = D(1)$, which happens to be the same as the original $O_1$ because the $Tdi$ function derives (based on the position relationship 1 < 2) that $O_1$ is performed on the left of $O_2$ in the linear state, hence its position is not affected by $O_2$.
3. Save $O_1'(t)$ in BUF = $[O_2(t), O_1'(t)]$.
4. Replay $O_1' = D(1)$ on "abce" to delete "b"; the document state becomes: "ace".

There is no need to store operations in the buffer indefinitely. As soon as *there is no future operation that could possibly be concurrent with the operations in the buffer* (a general garbage collection condition for OT) [56,68,85], those operations can be garbage collected and the buffer can be reset, i.e., BUF = [ ].

## 3.2 The CRDT Approach

*3.2.1 Key Ideas and Components.* WOOT [41,42] is commonly recognized as the first CRDT solution [49]. WOOT has two distinctive components. The first is a sequence of data objects, each of which is assigned with an *immutable* identifier and associated with either an existing character in the external document (visible to the user) or a deleted character (this internal object is then called a *tombstone*[7]). The second is the *identifier*-based operations, which are defined and applicable on the internal object sequence only.

For WOOT to work, an insert operation carries not only the identifier of the target object (i.e. the new character to be inserted), but also identifiers of two neighbouring objects corresponding to characters that are visible to the user at the time when the insert was generated. The target identifier and neighbouring object identifiers, together with tombstones embedded in the object sequence, are crucial elements in WOOT's solution to issues related to the FT puzzle [41,42].

Notwithstanding the existence of a variety of CRDTs, the life cycle of a user-generated operation in all CRDTs is essentially the same, and can be generally sketched as follows.

- When a local operation is generated by a user, it is immediately executed on the document visible to the user; then this operation is given as the input to the underlying CRDT solution. The CRDT solution converts the external *position*-based input operation into an internal *identifier*-based operation, applies the identifier-based operation to the internal object sequence, and propagates the identifier-based operation, to remote sites via a suitable external communication service.
- When a remote *identifier*-based operation is received from the network, the CRDT solution accepts it according to certain execution conditions [25,42], applies the accepted operation to the internal object sequence, and converts the identifier-based remote operation to a *position*-based operation, which is finally replayed on the external document state visible to the user at a remote site.

The above CRDT process of handling a user-generated operation (until replaying it at a remote site) naturally existed, but was often obscured in CRDT literature. We further elaborate this point under the general transformation framework in Section 4.

It should be noted that WOOT did not (and no other CRDTs did) actually change the formats of the external document state or operations, which are determined by the editing application [10]. For consistency maintenance purpose, WOOT (and other CRDTs) created an additional object sequence as an internal state, identifier-based operations as internal operations, and special schemes that convert between external and internal operations, search target objects or locations, and apply identifier-based operations in the internal state. See more discussions on the nature of CRDT internal object sequences and operations in Sections 4 and 5.

[7] To our knowledge, the AST (Address Space Transformation) solution in [19] was the first to use *marker* (like tombstone) objects to record deleted characters in co-editors.

*3.2.2. A Working Example for CRDT.* In Fig. 1-(c), we illustrate how the key functional components of WOOT work together to achieve the result in a simple scenario in Fig. 1-(a). This example also serves as an illustration of the general CRDT process sketched above.

At the start, each co-editing site is initialized with the same document state *"abe"* (visible to the user), and the same internal state (IS) consisting of a sequence of objects corresponding to the initial external document state:

*IS=<@s><a,ida,@s,idb,v><b,idb,ida,ide,v><e,ide,idb,@e,v><@e>*,

where *<@s>* and *<@e>* are two special objects marking the *start* and *end* points of an internal state; each of other objects has five attributes, e.g. *<b, idb, ida, ide, v>*, where *b* is the character represented by this object, *idb* is the identifier for this object, *ida* and *ide* are the identifiers for the two *neighboring* objects, respectively, and *v* indicates the character in this object is *visible* to the user (note: *iv* indicates the character is *invisible*). An object identifier is made of two integers (*sid, seq*), where *sid* is the identifier of the site that creates the object, *seq* is the sequence number of the operations generated at that site.

*Local Operation Handling*. User A interacts with the external document to generate a position-based operation $O_1 = D(1)$, resulting in a new state "ae". WOOT handles $O_1$ as follows:

1. Convert the position-based *D*(*1*) into the identifier-based *D*(*idb*) by:
   a. searching the object sequence, with the index position 1 in $O_1$, to locate the target object *<b,idb,ida,ide,v>* by counting only visible objects (*v* = true);
   b. creating an identifier-based *D*(*idb*), where *idb* is taken from *<b, idb, ida, ide, v>*.
2. Apply *D*(*idb*) to the object sequence by setting *iv* in the target object, which becomes a *tombstone* (also depicted by a line crossing the object in Fig. 1-(c)).
3. Propagate *D*(*idb*), rather than *D*(*1*), to User B.

Concurrently, User B interacts with the external document to generate a position-based operation $O_2 = I(2, \text{"}c\text{"})$, which results in a new state "*abce*". WOOT handles $O_2$ as follows:

1. Convert the position-based *I*(*2,"c"*) into the identifier-based *I*(*c,idc,idb,ide*) by:
   a. searching the object sequence, with the index position 2 in $O_2$, to find the two visible neighboring objects between the insert position in the object sequence by counting visible objects;
   b. creating an identifier-based operation *I*(*c, idc, idb, ide*), where *c* is the character to be inserted, *idc* is a new identifier for *c*, *idb* and *ide* are the identifiers of the two *neighboring* objects, respectively.
2. Apply *I*(*c, idc, idb, ide*) into the object sequence by creating a new object *<c,idc,idb,ide,v>* and injecting it at a *proper* location between the neighboring objects.
3. Propagate *I*(*c,idc,idb,ide*), rather than *I*(*2,* "*c*"), to User A.

*Communication and Operation Propagation:* The basic CRDT approach is independent of specific communication structures or protocols (more elaboration on this point later in this article). What is noteworthy here is that operations propagated under the CRDT approach are *identifier-based* operations defined on the internal object sequence, which is different from the OT approach.

*Remote Operation Handling*. At User B, the remote operation *D*(*idb*) is handled as follows:

1. Accept *D*(*idb*) for processing under certain conditions (e.g. the object to be deleted already exists in the object sequence) [25,42].
2. Apply *D*(*idb*) in the object sequence by:
   a. searching the object sequence, with the identifier *idb in D*(*idb*), to find the target *<b,idb,ida,ide,v>* with a matching identifier; and
   b. setting *iv* to the target object (to mark it as a tombstone).
3. Convert the identifier-based *D*(*idb*) into the position-based *D*(*1*), where the position parameter 1 is derived by counting the number of (visible) objects from the target object ~~*<b,idb,ida,ide,iv>*~~ to the start of the object sequence.
4. Replay *D*(*1*) on the external state to delete "b".

At User A, the remote operation *I*(*c, idc, idb, ide*) is handled as follows:

1. Accept *I*(*c,idc,idb,ide*) for processing under certain conditions [25,42].
2. Apply *I*(*c, idc, idb, ide*) in the object sequence by:
   a. searching the sequence, with identifiers *idb* and *ide* in *I*(*c, idc, idb, ide*), to find the two *neighboring* objects; and
   b. creating a new object *<c, idc, idb, ide,v>* and injecting it at a *proper* location between the two neighboring objects.
3. Convert the identifier-based operation *I*(*c, idc, idb, ide*) into a position-based operation *I*(*1, "c"*), where the position 1 is derived by counting the number of visible objects from the new object *<c, idc, idb, ide,v>* to the start of the object sequence.
4. Replay *I*(*1, "c"*) on the external state.

Finally, both sites reach the same final external and internal states. In WOOT and its variations (WOOTO [81] and WOOTH [4]), there exists no scheme to safely remove those tombstones. In some other tombstone-based CRDT solutions (e.g. RGA [46]), a garbage collection scheme was proposed to remove tombstones under certain conditions.

## 4 A GENERAL TRANSFORMATION FRAMEWORK

The concrete co-editing scenario (Fig. 1-(b) and (c)) is an instantiation of the general workflow of OT and CRDT solutions under a General Transformation (GT) framework for text co-editors. This GT framework is distilled in this work from a variety of OT, CRDT and other consistency maintenance solutions to co-editing. In this section, we first describe OT and CRDT under this GT framework, and based on these descriptions, we highlight the key characteristics and core components of the framework. Next, we examine OT and CRDT under this GT framework to reveal the major differences between *direct* transformation (OT) and *indirect* transformation (CRDT). Furthermore, we show the general applicability of the GT framework by describing another alternative solution − TTF (Tombstone Transformation Function) [43] under this framework, which reveals TTF's hybrid nature of CRDT and OT and clear up common misconceptions about TTF.

### 4.1 Key Characteristics and Core Components of the GT Framework

*4.1.1 Challenges and the Creation of the GT Framework.* One major challenge of this review and comparison work is the complexity and diversity of a large number of OT and CRDT solutions to be examined: they were designed in different algorithms, described using different and oftentimes obscured terminologies, with incomplete information and sometimes lacking critical technical details, and, worse yet, mixed with myths and flaws, which severely muddy the waters.

Drawn insights from experiences in designing and implementing numerous OT solutions, and dissecting a large number of CRDT solutions (and other alternative solutions), we have identified a set of basic functional components that are common to a wide range of consistency maintenance solutions for real-time co-editors. Based on those building blocks, we have developed this GT framework, which extracts high-level common functionalities of OT and CRDT solutions from their specific algorithmic details, and describes critical workflows in handling operations from its generation at the local site, all the way to its replay at a remote site, under both OT and CRDT.

We describe the working flows of OT and CRDT under the GT framework in Table 1. Based on this description, we elaborate the key characteristic and core components of the GT framework in the following subsections.

Table 1 Describing OT and CRDT under the GT framework. The shadowed blocks indicate common components shared by all transformation solutions for text editing.

| **The General Transformation (GT) Approach**<br>**Common external data and operation models, and consistency requirements**<br>**ES (External State)** is a sequence of characters: ES = $c_0, c_1, c_2, \ldots, c_n$.<br>**EO (External Operation)** is a position-based operation: EO = *insert*(*p*, *c*) or *delete*(*p*).<br>**Consistency requirements:** causality-preservation, convergence, and intention-preservation.<br>**General Transformation: GT($EO_{in}$)→$EO_{out}$: $EO_{in}$** is a user-generated input operation from a local document **$ES_{local}$**; **$EO_{out}$** is the output operation to be executed on a remote document **$ES_{remote}$**. | | |
|---|---|---|
| **Work Flow** | **OT** | **CRDT** |
| **Local User** | **User A** interacts with the local editor to generate a *position-based* **$EO_{in}$**, which takes effects on the **$ES_{local}$** immediately and then is given to the underlying LOH. | |
| **LOH** | **LOH($EO_{in}$) → $IO_t$:**<br>1. **Timestamp** a position-based **$EO_{in}$** to make $IO_t$.<br>2. **Save** $IO_t$ in the *operation buffer*.<br>3. **Propagate** $IO_t$ to remote sites. | **LOH($EO_{in}$) → $IO_{id}$:**<br>1. **Convert** a position-based **$EO_{in}$** into an identifier-based $IO_{id}$.<br>2. **Apply** $IO_{id}$ in the *object sequence*.<br>3. **Propagate** $IO_{id}$ to remote sites. |
| **CP** | $IO_t$ is *position-based* and defined on the external character sequence | $IO_{id}$ is *identifier-based* and defined on the internal object sequence |
| | A causally-ordered operation propagation and broadcasting service | |
| **ROH** | **ROH($O_t$) → $EO_{out}$:**<br>4. **Accept** a remote $O_t$ under certain conditions, e.g. *causally-ready*.<br>5. **Transform** $O_t$ against concurrent operations in the buffer to produce $O_t$' and **$EO_{out}$** (without a timestamp).<br>6. **Save** $O_t$ (and/or $O_t$') in the buffer.<br>7. **Replay** $EO_{out}$ on $ES_{remote}$. | **ROH($O_{id}$) → $EO_{out}$:**<br>4. **Accept** a remote $O_{id}$ under certain conditions, e.g. *causally-ready*.<br>5. **Apply** $O_{id}$ in the object sequence.<br>6. **Convert** identified-based $O_{id}$ back to position-based **$EO_{out}$**.<br>7. **Replay** $EO_{out}$ on $ES_{remote}$. |
| **Remote User** | **User B** observes the effect of the remote **$EO_{out}$** on **$ES_{remote}$**, which has the same effect of **$EO_{in}$** on **$ES_{local}$** observed by **User A**. | |

*4.1.2 External State and Operations versus Internal State and Operations.* One key characteristics of this framework is the explicit differentiation between the external document state and operations and the internal control state and operations.

The *external document state and operations* provide the common working context for all transformation-based consistency maintenance solutions:

- *External State* (ES) is accessible by the user for viewing and editing the document.
- *External Operation* (EO) is generated by the user for editing the document.

The nature and representation of the ES and EO are determined by the editing application, but independent of the underlying consistency maintenance solution, being OT or CRDT. In the domain of text editing, for example, ES represents a sequence of characters; EO represents one of two primitive operations *insert*(*p, c*) and *delete*(*p*), where *p* is a positional reference to the character sequence in ES, *c* is a character in ES (this parameter could be extended to a string of characters).

We should highlight that the modelling of the EO as *position*-based operations and the ES as *a sequence of characters* for text editors has been commonly adopted in existing consistency maintenance solutions, including OT and CRDT. This data and operation modelling is neither accidental nor merely a modelling convenience, but is consistent with and well-supported by decades of practice in building text editors [10,13,24,76]. The use of position-based operations does not imply the text sequence must be implemented as an array of characters. The positional reference to the text sequence has been implemented in numerous data structures, such as an array of characters, the linked-list structures, the buffer-gap structures, and piece tables [10,76].

On the other hand, the *internal state* and *operations* are created by the underlying system for consistency maintenance purpose:

- *Internal State* (IS) encapsulates all data structures for keeping track of consistency-impact information and is used internally only (invisible to the user).
- *Internal Operation* (IO) is converted from EO by the consistency maintenance solution and used internally only (invisible to the user).

Unlike ES and EO that are the same to all underlying consistency maintenance solutions, IS and IO representations are determined by individual solutions and may take a variety of forms. In OT, for example, the IS is represented as a buffer of operations, and the IO is a timestamped EO (both IO and EO are *position*-based). In CRDT, the IS is represented as a sequence of objects, which correspond to the sequence of characters in the EO, and (optionally) deleted characters (as tombstones); and the IO is based on immutable identifiers (i.e. *identifier*-based operations).

The differentiation of ES-EO from IS-IO is crucial to capture the *meta*-data-operation (IS-IO) used by individual solutions for consistency maintenance purpose, and helps to clear up misconceptions about CRDT object sequences and identifier-based operations (Section 5.2).

*4.1.3 End-to-End Description of the Full Life Cycle of User-Generated Operations.* Another distinctive characteristics of the GT framework is the end-to-end approach to describing consistence maintenance solutions, from local operation generation, handling, and propagation, all the way to remote operation acceptance, handling, and replay. This end-to-end approach covers the full life cycle of a user-generated operation in co-editors, and captures the big picture in which a consistency maintenance solution is operating. This approach is crucial to reveal important, but often hidden, similarities and differences among a variety of consistency maintenance solutions (shown in Table 1 and Table 2 in Section 4.3).

*4.1.4 Key Functional Steps and Core Components.* Key functional steps in the life cycle of a user-generated operation are covered under two core functional components of every transformation-based solution.

The first core component is *Local Operation Handler* (LOH), which encapsulates the data structures and algorithms for handling local operations, and covers the general steps below:

(1) *Converting* an external operation $EO_{in}$ (defined on $ES_{in}$) into an internal operation IO.
(2) *Integrating* IO to the internal state IS.
(3) *Propagating* the IO to remote sites, via an external communication and propagation (CP) service (another component in the framework to be explained below).

As shown in Table 1, OT and CRDT achieve above general Steps (1) and (2) differently:

- In OT, the conversion is achieved by *timestamping* $EO_{in}$ to make $IO_t$; the integration is achieved by *saving* $IO_t$ in the operation buffer.

- In CRDT, the conversion is achieved by *converting* $EO_{in}$ into an identified-based $IO_{id}$; and the integration is achieved by *applying* $IO_{id}$ in the internal object sequence.

The second core component is *Remote Operation Handler* (ROH), which encapsulates the data structures and algorithms for handling remote operations. ROH covers the general steps below:

(4) *Accepting* a remote IO from the external CP service, according to certain conditions (e.g. causality ordering [55] or execution conditions defined in [42]);
(5) *Converting* the remote IO to a suitable $EO_{out}$ (defined on $ES_{remote}$) according to consistency impact information recorded in the internal state.
(6) *Integrating* the remote IO's effect in the internal state.
(7) *Replaying* $EO_{out}$ on the remote $ES_{remote}$.

OT and CRDT achieve Steps (5) and (6) differently and in reverse orders: :

- In OT, the conversion is first performed by *transforming* a remote $IO_t$ with concurrent operations in the buffer to produce $EO_{out}$; the integration is then achieved by *saving* the $IO_t$ in the operation buffer.
- In CRDT, the integration is first performed by *applying* a remote $IO_{id}$ in the internal object sequence; the conversion from the $IO_{id}$ back to a position-based $EO_{out}$ is then achieved by searching and counting visible objects in the internal object sequence.

In addition to the core LOH and ROH components, the GT framework includes a *Communication and Propagation (CP)* component, which is responsible for broadcasting operations among co-editing sites via the network. Different transformation-based solutions may impose different conditions for propagating and accepting operations, and may or may not use a central server for any purposes related to co-editing. However, those differences are independent of whether or not the solution is OT or CRDT [75]. The inclusion of the CP component in the GT framework and the separation of CP from the core LOH and ROH components allow us to focus on the core transformation-related issues in LOH and ROH, without missing the communication factor in the big picture of the framework.

## 4.2 Examining OT and CRDT under the GT Framework

*4.2.1 Common Aspects of OT and CRDT.* As shown in Table 1, OT and CRDT share the same set of general consistency requirements [56]: *convergence*, *intention-preservation*, and *causality-preservation.* Also, they take the same *position*-based input operation $EO_{in}$ (defined on $ES_{local}$) at the local site, and produce a transformed *position*-based output operation $EO_{out}$ (defined on $ES_{remote}$) at a remote site.

Moreover, OT and CRDT share the same general requirement for the CP component: an external *causally-ordered* operation propagation and broadcasting service, which may or may not involve a central server (see detailed discussions on this point in [75]). One characteristic difference between OT and CRDT in relation to the CP component is: the propagated operations are *position*-based and defined on the external document state for OT solutions, but they are *identifier*-based and defined on the internal object sequence for CRDT solutions.

*4.2.2 Different Aspects of OT and CRDT.* Despite the above similarities, OT and CRDT differ significantly in the core components LOH and ROH. Fundamentally, every transformation-based consistency maintenance method must have a way to record the *concurrency-impact* information (as the internal state IS) arising from concurrent user actions, and to represent internal operations (IO) for consistency maintenance purposes. In OT solutions, concurrency-impact information is recorded in a buffer of concurrent operations; the IO is simply an external position-based operation with a timestamp. In CRDT solutions, concurrency-impact information is recorded in an internal object sequence, which maintains the effects of all (sequential or concurrent) operations applied to the document (plus those objects representing the initial characters in the document); and the IO is identifier-based, quite different from position-based external operations (EO). These differences are captured in the LOH component in Table 1.

For the ROH component, OT and CRDT use radically different methods to derive the transformed operation at a remote site. In OT solutions, when a remote position-based operation arrives, control algorithms process it against selected concurrent operations in the buffer one-by-one, and invoke transformation functions to do the transformation in each step. The actual transformation is based on a *compare-calculate* method, which compares numerical positions (using relations <, =, or >) between the two input operations, and calculates their positional differences (using arithmetic primitives + or −) to derive the new position of an output (transformed) operation, as illustrated in Fig. 1-(b).

In CRDT solutions (e.g. WOOT), when an identifier-based operation arrives at a remote site, it is first applied in the internal object sequence, then a position-based (transformed) operation is derived by using a *search-count* method, which searches objects in the sequence and counts the number of visible objects along the way, as illustrated for WOOT in Fig. 1-(c). Some CRDT solutions (e.g. Logoot [80,82]) do not use tombstones, so all their internal objects correspond to visible characters in the external state and the *search-count* method can be realized using binary-search, which is not the same as in WOOT, but with its own special issues, as specified in [80,82] and also discussed in detail in [74].

*4.2.3 Analysis of Compare-Calculate and Search-Count Methods.* In general, the arithmetic *compare-calculate* method adopted by OT is more efficient than the *search-count* method adopted by CRDT, as the former has a constant and low cost, but the latter has a variable and high cost that is dependent of editing positions (bounded by the document size).

To illustrate this point, we refer to the co-editing scenario in Fig. 1 again, in which there are two concurrent operations $O_1 = D(1)$ (to delete the character at position 1) and $O_2 = I(2, \text{“}c\text{”})$ (to insert *c* at position 2). Under any transformation approach, $O_2$ should be transformed into $O_2' = I(1, \text{”}c\text{”})$ (to compensate the left-shifting concurrency-impact effect of $O_1$), as shown in Fig. 1-(a). An OT solution can derive the position 1 in $O_2'$ with one subtraction, i.e. $1 = 2 - 1$, as illustrated in Fig. 1-(b). A CRDT solution (WOOT) can obtain the same result by searching and *counting* 2 objects and discounting one tombstone in the object sequence, as illustrated in Fig. 1-(c). In this trivial case, the WOOT cost is slightly higher than the OT cost, but not a big deal. However, what if $O_2 = I(p, \text{“}c\text{”})$, where $p$ has a value in a range from $10^3$ to $10^6$ (assuming the document size is in the same range as well, which is common for text documents)? An OT solution can get the transformation result $O_2' = I(p', \text{“}c\text{”})$, where $p' = p - 1$, by the same single subtraction as well. To derive this $p'$ value, however, WOOT has to search and count $p$ (ranging from $10^3$ to $10^6$) objects in the object sequence (and discount an arbitrary number of tombstones), which is multiple orders of magnitude more expensive than searching and counting 2 objects, in case that $O_2 = Insert(2, \text{“}c\text{”})$, and far more expensive than a single subtraction by OT!

*4.2.5 Direct versus Indirect Transformations.* To summarize, OT records the concurrency-impact information in a buffer of concurrent operations, and transforms position-based operations *directly* by selecting concurrent operations from the buffer, comparing and calculating positional differences between concurrent operations. In contrast, CRDT solutions record the effects of all (sequential and concurrent) operations in an internal object sequence, and transforms operations by: (1) converting a position-based operation into an identifier-based operation (and applying identifier-based operations in the object sequence) at a local site; and (2) converting an identifier-based operation back to a position-based operation at a remote site. Step (2) is achieved by applying identifier-based operations in the remote object sequence and then searching and counting visible objects in the internal object sequence. In other words, CRDT has adopted an *indirect* transformation approach, in contrast to the *direct* transformation approach taken by OT, to realizing the same general transformation. This direct versus indirect transformation difference has major impact on the correctness and complexity of OT and CRDT solutions, which are discussed in detail in [74].

### 4.3 Describing and Examining TTF under the GT Framework

The GT framework can be used to describe a wide range of consistency maintenance solutions, beyond OT and CRDT. To demonstrate this generality, we describe an alternative consistency maintenance solution TTF (*Tombstone Transformation Functions*) [43] under this framework, and reveal a hidden nature of TTF below.

TTF was proposed to solve the FT (False Tie) puzzle [54,56] in OT transformation functions for plain-text editing. TTF followed WOOT in using the idea of tombstone-based object sequences, and in fact, both of TTF and WOOT were proposed at nearly the same time by the same authors [41-43]. TTF was claimed to be the *first* and often cited as the *sole* correct OT solution [4,5,8,22, 41,43,46,82]. As such, TTF was often used as the OT representative in comparison with CRDT in CRDT literature. Quite some claims about CRDT superiority over OT were based on the comparison between CRDT solutions (e.g. Logoot [80,82], RGA [46]) and TTF (typically integrated with the SOCT2 algorithm [52]). For example, TTF was reported to be outperformed by Logoot and RGA for a factor up to 1000 in [4]. This 1000-times-gain claim was widely cited as an experimental evidence for CRDT's performance superiority over OT (e.g. [4,5,8,46,81]).

Validating whether and how CRDT solutions (e.g. Logoot and RGA) had *truly* achieved 1000-time-gain over TTF (+SOCT2) would be interesting, but outside the scope of this paper. What we want to point out here is that those CRDT and TTF claims are actually *groundless* and *false,* because: (1) they are contradicted by the facts that numerous OT solutions had been proven to be correct, with respect to well-established conditions and properties, *before* and *after* TTF and WOOT appeared (see [74] for comprehensive and critical review of OT and CRDT in correctness and complexity); and (2) they are also mistaken about what TTF really is: TTF is in fact a *hybrid* of CRDT and OT, or "*a CRDT in disguise*" (see footnote 4).

We describe TTF under the GT framework in Table 2. The LOH component of TTF takes, as input, a user-generated operation $\mathbf{EO_{in}}$, which is position-based and defined on the local external state $\mathbf{ES_{local}}$, and produces, as output, the internal operation $\mathbf{IO_{i,t}}$, which is still position-based and has a timestamp (like OT), but defined on the internal tombstone-based object sequence (like CRDT). Internal Steps (1) and (2) in LOH of TTF in Table 2 are the *union* of corresponding steps for OT and CRDT in Table 1, which shows that TTF LOH is a mixture of the OT and CRDT LOH components. TTF propagates the internal operation $\mathbf{IO_{i,t}}$, which is a position-based operation (like OT), but defined on the internal tombstone-based object sequence (like CRDT).

The ROH component of TTF takes, as input, a remote operation $\mathbf{IO_{i,t}}$, which is defined on the internal tombstone-based object sequence, and produces, as output, the external operation $\mathbf{EO_{out}}$, which is *position-based* and defined on the remote external state $\mathbf{ES_{remote}}$. Similarly, internal Steps (5) and (6) in ROH of TTF in Table 2 are the union of the corresponding steps in ROH for OT and CRDT in Table 1, which shows that the TTF ROH is a mixture of OT and CRDT ROH components. Due to its hybrid nature, TTF bears the costs of both CRDT and OT, with the main costs dominated by its CRDT components (in Steps 1-(a) and 2-(a) in LOH and Step 6-(c) and -(d) in ROH in Table 2). We refrain from detailed comparison of TTF with OT or CRDT in this paper, but plan to do comprehensive comparisons of OT, CRDT, TTF, and other alternatives (e.g. [19,28-31]), which can all be described under the GT framework in a future paper.

Table 2. Describing TTF (a hybrid OT and CRDT) under the GT framework.

| **The General Transformation (GT) Approach**<br>**Common external data and operation models, and consistency requirements**<br>**(the same as in Table 1)** | |
|---|---|
| **Work Flow** | **TTF (CRDT+OT)** |
| **Local User** | **User A** interacts with the local editor to generate a *position-based* **$EO_{in}$**, which takes effects on the **$ES_{local}$** immediately and is given to the underlying LOH. |
| **LOH** | **LOH($EO_{in}$) → $IO^i_t$ :**<br>1. **Convert $EO_{in}$** to **$IO_{i,t}$**:<br>a. **Convert** an external position-based **$EO_{in}$** into an internal position-based **$IO_i$** based on the internal object sequence with tombstones (**CRDT**).<br>b. **Timestamp** $IO_i$ to make $IO_{i,t}$ (**OT**).<br>2. **Integrate $IO_{i,t}$** in the internal state**:**<br>a. **Apply $IO_i$** in the internal object sequence (**CRDT**).<br>b. **Save $IO_{i,t}$** in the operation buffer (**OT**).<br>3. **Propagate $IO_{i,t}$** to remote sites (**CRDT/OT**). |
| **CP** | **$IO_{i,t}$** is *position-based* (like **OT**), but defined on *an internal object sequence* (like **CRDT**) |
| | A causally-ordered operation propagation and broadcasting service (CRDT/OT) |
| **ROH** | **ROH( $IO_{i,t}$) → $EO_{out}$:**<br>4. **Accept** a remote **$IO_{i,t}$** under certain conditions, e.g. *causally-ready* (**OT/CRDT**).<br>5. **Convert $IO_{i,t}$** to **$EO_{out}$**, and<br>6. **Integrate $IO_{i,t}'$** to the internal state are collectively achieved as follows:<br>a. **Transform $IO_{i,t}$** against concurrent operations in the buffer to produce **$IO_{i,t}'$** (**OT**);<br>b. **Save $IO_{i,t}'$** (and/or **$IO_{i,t}$**) in the operation buffer (**OT**);<br>c. **Apply $IO_{i,t}'$** in the internal object sequence (**CRDT);**<br>d. **Convert $IO_{i,t}'$** (defined on the internal object sequence) to **$EO_{out}$** (defined on the remote external state **$ES_{remote}$**) by searching the internal object sequence (**CRDT**).<br>7. **Replay $EO_{out}$** in **$ES_{remote}$** (**CRDT/OT**). |
| **Remote User** | **User B** observes the effect of the remote **$EO_{out}$** on **$ES_{remote}$**, which has the same effect of **$EO_{in}$** on **$ES_{local}$** observed by **User A**. |

## 5 DISCUSSIONS

With the guidance of the GT framework, we further explore what CRDT really *is* and *is not* for co-editors in this section.

### 5.1 The Hidden Transformation Nature of CRDT

When we put OT and WOOT solutions to the same co-editing example side-by-side in Fig. 1, it is clear that both solutions produce identical position-based operations, i.e. $O_2 = I(2, \text{"c"})$ is transformed into $O_2' = I(1, \text{"c"})$ at User A, while $O_1 = D(1)$ is unchanged at User B. The reader can verify this by comparing Fig. 1-(c) (for WOOT) and Fig. 1-(b) (for OT). This is an intuitive example that shows WOOT indeed is an alternative to realizing the general transformation.

More generally, when we examine the CRDT approach under the GT framework in Table 1, the transformation nature of CRDT becomes clear as well: CRDT and OT take the same position-based input operation $EO_{in}$ (defined on $ES_{local}$) at the local site, and produce a transformed position-based output operation $EO_{out}$ (defined on $ES_{remote}$) at a remote site.

Why was the transformation nature of CRDT unknown previously? We draw attention to Steps 3 and 4 in handling a remote operation described in Section 3.2.2 for the scenario in Fig. 1-(c) (or Steps 6 and 7 in ROH for CRDT in Table 1). These two steps play the role in converting an identifier-based operation into a position-based operation, and replaying a position-based operation on the external state to ensure consistency. However, both steps were omitted in the description of WOOT [42] and its variations: the final step of handling a remote operation ends at Step 2 in handling a remote operation in Section 3.2.2 (or Step 5 in ROH for CRDT in Table 1), i.e. after integrating the identifier-based operation into the internal object sequence. For the scenario in Fig. 1-(c), if Steps 3 and 4 were skipped, User B would still see the document as *"abce"* even after the remote operation *D*(*idb*) has been integrated into the internal object sequence, while User A would continue to see the document as *"ae"* after *I*(*c, idc, idb, ide*) has been internally processed. In each case, the external documents visible to Users A and B are neither convergent nor intention preserving. It is clear that these steps are not mere implementation details, but crucial steps to ensure the correctness of a consistency maintenance solution for co-editing.

In WOOT [42] , a *value(S)* function was briefly mentioned and supposed to map the internal object sequence *S* to the external state visible to the user. However, there was no hint on when and how the *value(S)* function might be invoked to map the internal object sequence *S*, to accomplish the final effect of replaying a remote operation on the external document. For the sake of correctness and real-time update of the external document, *value(S)* should be invoked whenever a remote identifier-based operation is integrated into the internal object sequence. In principle, the *value(S)* function could be implemented in two alternative ways. One is to derive a position-based operation and apply this operation to the external document, which is what was illustrated in Fig. 1-(c). The other is to: (1) scan the internal object sequence to extract visible characters and generate a new sequence by character-wise concatenation, and (2) reset the external document state with the generated sequence of characters, which will include the effect of the newly integrated remote operation. The second alternative is generally more expensive than the first one. One way or another, handling a remote operation must include the steps that change the external document visible to the user.

Furthermore, we found those missing steps in CRDT publications manifested themselves in the documentation and/or source code of co-editors based on CRDTs (e.g. WOOT [41,42] and Logoot [80,82]), which were built by practitioners who were interested in learning whether and how CRDTs actually work in real editing environments [75]. In prototyping those co-editors, some practitioners also detected other missing "*key details on how to handle certain edge cases*" (see footnote 10 in [74]) and various *abnormalities*[8] (see discussions in Section 4.4. in [74]), which were actually symptoms of deep algorithmic flaws in published CRDTs. Unfortunately, none of those prototype co-editors was built by researchers who published theoretic CRDTs, and the discoveries (or feedback) from building those co-editors by practitioners had little impact on follow-up CRDT research [74]. Since the start (WOOT), CRDT research has adopted predominantly theoretic approaches to identifying co-editing issues, designing and verifying solutions (e.g. using theorem provers, model checkers, or mathematic proofs) [7,22,23,41,42,43], but rarely implemented CRDT solutions in working co-editors for experimental validation. Consequently, theoretic CRDT work missed not only some crucial steps in co-editing (which masked the transformation nature of CRDT), but also (and more critically) the big picture of a co-editing system, and hence failed to learn (or chose to ignore) the hidden CRDT correctness and complexity issues, which are examined in detail in [74,75].

### 5.2 Demystifying the Commutativity Property of CRDT

As highlighted in Section 4.1, the GT framework possesses two distinctive characteristics: one is the clear differentiation of ES-EO (used in external editors and visible to users) from IS-IO (used by an underlying consistency maintenance solution); and the other is the end-to-end coverage of the full life cycle of user-generated operations in real-time co-editors. With the guidance of these two points in the GT framework, we reveal common misconceptions about CRDT object sequences and operations and demystify the commutativity property of CRDT below.

[8] https://stackoverflow.com/questions/45722742/logoot-crdt-interleaving-of-data-on-concurrent-edits-to-the-same-spot.

One common misconception about CRDT object sequences and identifier-based operations is that they are *native* to the editor. This misconception leads to the illusion that there is no need for position-based operations in CRDT, let alone the need to convert them to/from identifier-based operations. Evidences from existing CRDTs suggest otherwise: for tombstone-based WOOT variations [4,41,42,81] and RGA [46], the conversion of *local* position-based operations into identifier-based operations was explicitly described in publications, although the conversion of *remote* identifier-based operations back to position-based operations was omitted; for non-tombstone-based solutions, such as Logoot variations [80,82], the conversion of *remote* identifier-based operations back to position-based operations was explicitly described, but the conversion of *local* position-based operations into identifier-based operations was obscured. Clearly, designers of these CRDT solutions were cognizant of the fact that CRDT identifier-based operations and object sequences were invented for CRDT scheme descriptions, but not native to text editors. Furthermore, in all existing CRDT-based co-editors (see [75]), CRDT object sequences and identifier-based operations are not native but external to real editors. The root of this misconception is the confusion of CRDT internal object sequences and identified-based operations (i.e. IS-IO under the GT framework) with the external document states and operations used by the editor (i.e. ES-EO under the GT framework), which is common in CRDT literature.

While it is unquestionable there exists no single CRDT that is native to any existing editor, some may still argue for a *possibility* that CRDT object sequences and identifier-based operations might be adopted in future editors for co-editing. Unfortunately, experiences and insights from past co-editing research and practice suggest that CRDT object sequences and identifier-based operations are poor candidates to be considered for use as native data structures and operations for text (or other) editors, and by deduction for co-editors, for the following reasons.

1. Data structures and operations of text editors ought to be designed for effective and efficient support for standard text editing operations and user interactions. There exists substantial well-established prior art on how to create and optimize text editors that are performant (e.g. initial loading time, memory paging speeds, etc.) [10,13,24,76] – desirable properties that should be preserved in co-editors as well. However, CRDT object sequences and identifier-based operations were invented for supporting CRDT-based consistency maintenance, without any concern for efficient support of standard text editing functionalities in their makings.
2. Existing research has found that published CRDT object sequences, operations and manipulation schemes have high time and space complexities and various correctness issues, as discussed in detail in [74], for serving the intended consistency maintenance purpose; it is inconceivable to use them as the basis for supporting unintended conventional editing functions in standard text (or other) editors.
3. Last but not least, past co-editing research and practice in building real world co-editors suggested that co-editors ought to be built by *separating*, rather than *mixing*, concerns about consistency maintenance from concerns for conventional editing functions, to allow for modularity, simplicity, and efficiency of both conventional editing functions and consistency maintenance solutions (see detailed discussions in [75]). In OT-based co-editors [1,56,58,61,62,69,83], for example, the choice of strategies (e.g. what native data structures or operation models to use) for implementing efficient document editing is completely left to application designers, and the support for real-time collaboration is orthogonal to and interfaced with the editing application by exposed abstract-data-type, which is, in the case of text editing, a sequence of characters [10,24]. The idea to mix data structures and operations devised for consistency maintenance (e.g. CRDT object sequences and identifier-based operations) within editors is not supported, but contradicted, by experiences and insights from co-editing research and real world applications (see [75]) and practices[9].

Closely related to the above misconception is the notion that CRDT makes concurrent operations *natively* commutative or *by design*, whereas OT makes concurrent operations commutative *after the fact* [48,49]. The fallacy of this notion is its confusing the commutativity of *identifier*-based operations in the internal CRDT object sequence with the commutativity of *position*-based operations on the external text document visible to users. The fact is, as revealed above, CRDT identifier-based operations are not native to editors, but only used within the CRDT object sequence, and have to be converted from/to position-based operations in order to make them commutative in the document visible to users. In contrast, OT solutions directly transform concurrent position-based operations to make them commutative on the text sequence visible to users. CRDT has to achieve the same OT commutativity *after the fact* as well, albeit *indirectly,* as revealed by the end-to-end description of CRDT under the GT framework.

Uncovering the transformation nature and demystifying the commutativity property of CRDT are crucial in detecting and understanding the real differences between OT and CRDT in achieving the same commutativity of position-based operations on the external text document visible to users – the real objective of consistency maintenance for co-editors. Unfortunately, the CRDT way of achieving this objective turned out to be highly complex and error-prone, which is examined in details in [74].

### 5.3 General Differences between OT and CRDT in Time and Space Costs

While both OT and CRDT have followed the same GT approach to co-editing, they have taken radically different (*direct* vs *indirect*) approaches to realizing this GT. Particularly, they have adopted different strategies to record the concurrency impact information – *an internal operation buffer* (for OT) versus *an internal object sequence* (for CRDT), which have had fundamental impacts on the design and complexity of OT and CRDT solutions.

Without diving into algorithmic details of specific OT or CRDT solutions, we highlight some general and characteristic differences between OT and CRDT in complexities and costs below:

1. **Variables in Determining Time and Space Complexities.** As OT records the concurrency impact information in an internal operation buffer, the time and space complexity of an OT solution depends on a variable $c$ (for *concurrency*) – the number of operations saved in the buffer and involved in transforming an operation. The value of $c$ is related to concurrency but unrelated to the document contents; and $c$ is often bounded by a small value, e.g. $0 \leq c \leq 10$, for real-time sessions with a few (< 5) users. In contrast, CRDT uses an internal object sequence to record the concurrency impact information, the time and space complexity of a CRDT solution depends on a variable $C$ (for *Contents*) or $C_t$ (for *Content with tombstones*) – the number of objects in the internal object sequence. The value of $C/C_t$ is determined by the document contents but unrelated to concurrency; and $C$ is typically orders of magnitude larger than $c$, e.g. $10^3 \leq C \leq 10^6$, for common text document sizes ranging from 1K to 1M characters, while $C_t$ is much larger than $C$ owing to the inclusion of tombstones. In real-time text co-editing, the following *inequality* commonly holds: $C_t \gg C \gg c$, which have major impacts on the theoretic complexity and practical efficiency of OT and CRDT solutions. In general, CRDT solutions have significantly higher time and space complexities (determined by $C_t$ and $C$) than OT solutions (determined by $c$), as revealed in [74].

[9] "*Why CRDT didn't work out as well for collaborative editing xi-editor*", https://news.ycombinator.com/item?id=19886883. This blog hosts discussions on some issues and lessons from an unsuccessful attempt in using CRDT to build a text editor *xi*.

2. **Co-Editing Session Initialization.** At the start of a co-editing session, the operation buffer for an OT solution is empty, bearing no space and time cost in initialization. In contrast, the internal object sequence for a CRDT solution must be created to represent initial characters in the document[10], which incurs space and time overhead at the initialization time and bears the cost during a whole session. Session initialization complexity can make big differences in session management and handling later-comers during a co-editing session [6,61].
3. **Handling Sequential and Concurrent Operations.** An OT solution has no time and space cost for transformation when there is no concurrent operation (with $c = 0$) as the operation buffer can be emptied with garbage collection[11]. In contrast, a CRDT solution bears similar time and space costs regardless whether operations are sequential or concurrent as all operations must be applied in the internal object sequence (with costs determined by $C$ or $C_t$), which can never be emptied[12] unless the document itself is empty.
4. **Handling Local and Remote Operations.** An OT solution has no transformation cost in handling local operations since a local operation can never be concurrent with any operation in the buffer. In contrast, a CRDT solution bears nearly the same processing costs regardless whether an operation is local or remote since every operation has to be applied in the internal object sequence. The longer time the local operation processing takes, the less responsive the co-editor is to the local user.

Readers are referred to [74] for detailed comparison in time and space complexity, as well as correctness, among representative OT and CRDT solutions.

## 6 CONCLUSIONS

In this work, we have conducted comprehensive and critical reviews of OT and CRDT solutions for consistency maintenance in real-time co-editing, and made a number of important discoveries, which contribute to the state-of-the-art knowledge on collaboration-enabling technology in general, and on OT and CRDT in particular.

In this paper, we have presented a novel general transformation (GT) framework, which provides a common ground for describing, examining and comparing a variety of consistency maintenance solutions in co-editing, including OT and CRDT solutions, among others. The GT framework has two key characteristics: one is the explicit differentiation of external document states and operations (used by external editors and visible to users) from internal control states and operations (used by underlying consistency maintenance techniques); and the other is the end-to-end coverage of the full life cycle of user-generated operations in real-time co-editing sessions. The GT framework contains two core components, i.e. LOH (Local Operation Handler) and ROH (Remote Operation Handler), for capturing general steps and common functions in processing operations at local and remote co-editing sites in transformation-based solutions. These key characteristics and core components are distinctive and collectively make the GT framework unique in its capability of describing common functionalities of a variety of consistency maintenance algorithms. This GT framework has played a crucial role in guiding us to detect pitfalls in existing consistency maintenance solutions, and can also be used to guide people to design new consistency maintenance solutions for co-editors, and avoid trapping in similar pitfalls, in the future.

---

[10] Nearly all CRDT articles ignored the existence and impact of initial document contents in calculating the size of the internal object sequence (see detailed analysis in [74]).

[11] Operation garbage collection is commonly used in OT and OT-based co-editors, e.g. [12,37,54,56,61,62,68,85].

[12] Tombstones can be removed as garbage in some tombstone-based CRDT solutions (e.g. RGA[46]), but not in others (e.g. WOOT variations [4,41,42,81]). However, tombstone collection does not address the *object sequence overhead* issue, which is the CRDT-special overhead and exists in all CRDT solutions [74].

The GT framework has served as a powerful lens for us to examine and make important discoveries about OT and CRDT. In particular, we have revealed hidden but critical facts about CRDT: CRDT is like OT in following the same GT approach to consistency maintenance in real-time co-editors; CRDT is the same as OT in making user-generated operations commutative *after the fact*, albeit *indirectly*; and CRDT operations are not natively commutative to text editors, but require additional conversions between CRDT internal operations and external editing operations. Revealing these facts helps demystify what CRDT really *is* and *is not* to co-editing, which in turn helps bring out the real differences between OT and CRDT – their radically different ways of realizing the same GT approach and achieving the same commutativity for co-editors.

Without diving into algorithmic details of specific solutions, we have outlined in this paper some general differences between OT and CRDT determined by their basic approaches, which include the time and space complexities and costs in handling concurrent and sequential operations, handling local and remote operations, and initializing co-editing sessions. Last but not least, we have revealed the different natures of the complexity variables for OT and CRDT. OT time and space complexities depend on a variable $c$ (for *concurrency*) – the number of concurrent operations involved in transforming an operation. In contrast, CRDT time and space complexities depend on a variable $C$ (for *Contents*) or $C_t$ (for *Content with tombstones*) – the number of objects maintained in the internal object sequence. In real-time text co-editing, the following *inequality* commonly holds: $\boldsymbol{C_t \gg C \gg c}$, which have major impacts on the theoretic complexity and practical efficiency of OT and CRDT solutions.

The GT framework and the discoveries based on this framework provide the foundation to reveal real differences between OT and CRDT for co-editors in correctness and complexity, as well as in building real world co-editors, which are reported in follow-up papers [74,75]. Our discoveries from this work revealed facts and evidences that refute CRDT superiority claims over OT on all accounts, which helps to explain the underlying reasons behind the choices between OT and CRDT in the real world.

Past co-editing research and development have explored various alternative consistency maintenance solutions, and accumulated a wealth of experiences and lessons. The time is ripe to review those alternatives critically and to learn what each major alternative really is, whether or not it works and why, and whither it is heading. For any alternative to be a viable solution in co-editing, in our view, it should be able to offer capabilities that are *truly* superior to existing state-of-the-art solutions, and demonstrate the relevance in supporting real world co-editors.

We hope discoveries from this work will help clear up common myths and misconceptions surrounding OT and CRDT in co-editing literature, inspire fruitful explorations of novel collaboration techniques, and accelerate progress in co-editing technology innovation and real world applications.

## ACKNOWLEDGMENTS

This research is partially supported by Academic Research Fund Tier 2 (MOE2015-T2-1-087) Grant from Ministry of Education, Singapore. The authors are grateful to the anonymous expert reviewers for their insightful and constructive comments and suggestions, which have helped improve the presentation of this article.